\documentclass[twocolumn,showpacs,groupedaddress,assymb,amsmath]{revtex4}
\usepackage{graphicx}
\usepackage{mathrsfs}
\usepackage{amssymb}
\usepackage{amsmath}
\usepackage{extarrows}

\begin{document}

\title{Robust state transfer with high fidelity in spin-1/2 chains by Lyapunov control}

\author{Z. C. Shi$^{1}$, X. L. Zhao$^{1}$,
and X. X. Yi$^{2}$\footnote{To whom correspondence should be
addressed: yixx@nenu.edu.cn}}
\affiliation{$^1$ School of Physics and Optoelectronic Technology\\
Dalian University of Technology, Dalian 116024 China\\
$^2$ Center for Quantum Sciences and School of Physics, Northeast
Normal University, Changchun 130024, China }


\begin{abstract}
Based on the  Lyapunov control,   we present  a scheme  to realize
state transfer with high fidelity by only modulating   the boundary
spins in a quantum spin-1/2 chain. Recall that the
conventional transmission protocols aim at  non-stationary  state
(or information) transfer  from the first spin to the end spin at a
fixed time. The present scheme possesses the following advantages.
First, the scheme does not require  precise manipulations of the
control time. Second, it is robust against uncertainties in the
initial states and fluctuations in the control fields. Third, the
controls are exerted only on the boundary sites of the chain. It
works for variable spin-1/2 chains with different periodic
structures and has well scalability. The feasibility to replace the
control fields by square pules is explored, which simplifies the
realization in experiments.
\end{abstract}

 \maketitle

\section{introduction}

Quantum information processing (QIP) has been extensively studied in
the past decades.  One of the main challenges in actual physical
implementations  of QIP is   to transfer quantum information between
different elements in quantum networks. In this respect, spin chain
systems  as quantum channels\cite{bose07,kay10,apollaro13} to
transfer   information play an important role in QIP. It is believed
that almost all spins would participate in the dynamics of an
unmodulated spin chain \cite{bose03} due to spin-spin couplings,
leading to dispersions of the chain  that are harmful for quantum
state transfer. Recently, several strategies have been proposed to
avoid such dispersions (see, e.g.,
\cite{christandl04,fitzsimons06,venuti07,gualdi08,
feldman09,banchi10,ping13,zwich13,banchi11}). The first approach is
mainly based on modulating coupling strengths between nearest
neighbors to reduce the effect of dispersion
\cite{wang11,bruderer12,vinet12}. The second  depends on the weak
interaction of boundary spins to the remainders (bulk spins), making
the bulk spins almost un-excited during the time evolution. The
dynamics in the second method can be viewed as an effective Rabi
oscillation between the boundary spins
\cite{bruderer12,paganelli06,wojcik07,venuti07pra}. In the other
sort of schemes, such as modulating the Larmor frequencies on sites
\cite{feldman10} or adding external potentials \cite{giorgi13} to
the spin chain,  perfect state transfer can be obtained only for a
fixed time. This means that a precise control on the evolution time
is strictly required to obtain high fidelity state transfer. Once
there exist errors in the coherent time evolution, the fidelity
decreases sharply.

One important progress to take the disorders into account was made
in Ref. \cite{yao11}, where an robust state  transmission in random
unpolarized spin chains was proposed. This proposal  is immune to
some significant types of disorders and does not need to manipulate
individual spins. In contrast, by manipulating the individual
spin-spin couplings, Zwick and co-workers \cite{zwick11,zwick12}
investigated the performance for quantum state transfer and showed
its robustness against static perturbations.

Lyapunov-based control technique has attracted  many attentions due
to its powerful applications  in   manipulating quantum states in
quantum  systems
\cite{alessandro07,beauchard07,kuang08,coron09,wang09,wang10,hou12,yi09}.
To apply the  Lyapunov control, a function of  the target state
called Lyapunov function $V$ has to be  specified, where the control
field $f(t)$ is designed to guarantee the Lyapunov function
decreasing monotonously, i.e., $\dot{V}\leq0$. The merit of the
Lyapunov control is that the target is asymptotically approached,
which means the manipulation time is not important. In this paper,
we apply the Lyapunov-based control technique to the  quantum state
transfer  from one boundary spin to another. We show that  by
manipulating the interaction strengths between the boundary spins
and the bulk spins in an one-dimensional spin-1/2 chain, the quantum
state transfer can be realized. To show the robustness of this
control, we numerically calculate the   performance under the effect
of external perturbations.

The paper is organized as follows. In section \ref{II}, we present a
general formalism for state transfer in a spin-1/2 chain by Lyapunov
control. In section \ref{III}, we select a specific eigenstate of
the spin-1/2 chain for state transfer. The control fields are
designed and the robustness against fluctuations is investigated in
section \ref{IV}. The scalability of the state transfer in such a
spin-1/2 chain and the improvement on the control fields are briefly
discussed in section \ref{V}. Finally, we give a conclusion in
section \ref{VI}.

\section{general formalism for  state transfer in spin-1/2 chain}\label{II}

We start by considering  an one-dimensional spin-1/2 chain with
modulated XY interactions between nearest neighbors. The free
Hamiltonian can be written as,
\begin{eqnarray}\label{1}
H_{0}^{xy}=\frac{1}{2}\sum_{m=1}^{M-1}D_{m}(S_{m}^{x}S_{m+1}^{x}
+S_{m}^{y}S_{m+1}^{y})-\frac{1}{2} \sum_{m=1}^{M}\Omega_mS_m^z,
\end{eqnarray}
where $\Omega_m$ is the Larmor frequency and $D_{m}$ denotes the
coupling strength between the $m$-th and $(m+1)$-th spins. $S_m^i
(i=x,y,z)$ is the Pauli matrix and $M$ is the length of spin-1/2
chain. We label the first spin and the end spin by $1$ and $M$,
respectively. Clearly, $[H_{0}^{xy},S^{z}]=0$ where
$S^{z}=\sum_{m=1}^{M}S_{m}^{z}$, means that the total number of spin
up (down) is conserved. When quantum information is encoded on the
state of spin up and spin down in the spin-1/2 chain, the state
transfer can be equivalent to information transfer from the first
spin to the end spin.

By Jordan-Wigner transformation, the spin Hamiltonian $H_{0}^{xy}$
can be mapped into a spinless fermions Hamiltonian,
\begin{eqnarray}\label{15}
H_{0}=\sum_{m=1}^{M-1}D_{m}(c_{m}^{\dag}c_{m+1}+c_{m+1}^{\dag}c_{m})+
\sum_{m=1}^{M}\Omega_mc_m^{\dag}c_m,
\end{eqnarray}
where $c_m$ ($c_m^{\dag}$) represents the annihilation (creation)
operator of spinless fermion at site $m$. Due to
$[H_0,\sum_{m=1}^{M}c_m^{\dag}c_m]=0$, one can decompose the Hilbert
space $\mathcal{H}$ into subspaces
$\mathcal{H}=\sum_j\mathcal{H}_j$, each of them such as
$\mathcal{H}_j$, possessing  a fixed number of spinless fermions
$j$. We then choose the subspace $\mathcal{H}_0$ and
$\mathcal{H}_{1}$ for the state transfer, in which it has one
excitation at most. Apparently, the dimension of subspace
$\mathcal{H}_{1}$ is equal to the number of spins in the chain and
we denote the state with a single excitation at site $m$ by
$|\textbf{m}\rangle$. The Hamiltonian $H_0$ can then be written in a
matrix form in the basis $\{|\textbf{m}\rangle\}$,
\begin{equation}\label{3}
{H_0}=\left( \begin{array}{ccccccc}
   \Omega_1 & D_1 & 0 & 0 & \cdots & 0 & 0  \\
   D_1 & \Omega_2 & D_2 & 0 & \cdots & 0 & 0  \\
   0 & D_2 & \Omega_3 & D_3 & \cdots & 0 & 0 \\
   0 & 0 & D_3 & \Omega_4 & \cdots & 0 & 0 \\
   \vdots & \vdots & \vdots & \vdots & \ddots & \vdots & \vdots \\
   0 & 0 & 0 & 0 & \cdots & \Omega_{M-1} & D_{m-1} \\
   0 & 0 & 0 & 0 & \cdots & D_{M-1} & \Omega_M \\
\end{array} \right),
\end{equation}
where the vacuum state $|\textbf{0}\rangle=|00\cdots0\rangle$, and
$|\textbf{m}\rangle=c_{m}^{\dag}|\textbf{0}\rangle=|0\cdots01_m0\cdots0\rangle$
\cite{note1}. In the following, we study the problem of state
transfer by using this matrix form of system Hamiltonian.

To be specific, the state transfer can be obtained by free evolution
of the Hamiltonian $H_0$
\cite{bose03,christandl04,bruderer12,feldman10}. It can be
formulated as follows: The initial state of the first spin is
$\alpha|0\rangle+\beta|1\rangle$, while the remaining spins are in
spin down state,
\begin{eqnarray}\label{15}
|\psi(0)\rangle&=&\alpha|0_{1}0\cdot\cdot\cdot00_{M}\rangle+
\beta|1_{1}0\cdot\cdot\cdot00_{M}\rangle\nonumber\\
&=&\alpha|\textbf{0}\rangle+\beta|\textbf{1}\rangle,\label{ini1}
\end{eqnarray}
where $|\alpha|^2+|\beta|^2=1$.  After a fixed  evolution time
$t=T$, the system would evolve to,
\begin{eqnarray}\label{M}
|\psi(T)\rangle&=&\alpha|0_{1}0\cdot\cdot\cdot00_{M}\rangle
+\beta|0_{1}0\cdot\cdot\cdot01_{M}\rangle\nonumber\\
&=&\alpha|\textbf{0}\rangle+\beta|\textbf{M}\rangle.
\end{eqnarray}
Those proposals  need to control the evolution time precisely to get
perfect state transfer. Especially, the final state in those
proposals is not a steady state. In the following, by Lyapunov
control, we show that in the spin-1/2 chain  one can realize steady
state transfer with high fidelity. The  control process can be
illustrated as,
\begin{eqnarray}\label{15}
|\psi(0)\rangle=\alpha|\textbf{0}\rangle+\beta|\textbf{1}\rangle
\xLongrightarrow[\texttt{control}]{\texttt{Lyapunov}}|\psi(T)\rangle
=\alpha|\textbf{0}\rangle+\beta|\lambda_f\rangle,
\end{eqnarray}
where $|\lambda_f\rangle$ denotes  the target state and one of
eigenstates of Hamiltonian $H_0$ at the same time. Because
$|\textbf{0}\rangle$ is the ground state of spin-1/2 chain, and we
choose the control Hamiltonian such that the ground state remains
unchange in the dynamics, the state transfer mainly focuses on
steering the initial state $|\textbf{1}\rangle$ into the target
state $|\lambda_f\rangle$.

We should point out  that the quantum information  encoded
in the first spin  is almost perfectly transferred to the end spin
in our system, like  the other protocols
\cite{bose03,christandl04,fitzsimons06,venuti07,
gualdi08,feldman09,banchi10,ping13,zwich13,banchi11} did for state
transfer. The difference is that   our resulting  state is the
eigenstate $|\lambda_f\rangle$ rather than the state
$|\textbf{M}\rangle$ in equation (\ref{M}). This change has many
advantages as we mentioned in the abstract, the price we have to pay
is  the state (or information) is not localized well at the end
spin, leading to imperfect state transfer. This   can be improved by
designing the Hamiltonain as we show below.

When adding control fields $f_k(t)$  with control Hamiltonians $H_k$
into the system, the time evolution of spin-1/2 chain
  satisfies  the following Schr\"odinger equation ($\hbar=1$):
\begin{eqnarray}\label{7}
|\dot{\psi}\rangle=-i\Big[H_0+\sum_{k=1}^{K}f_k(t)H_k\Big]|\psi\rangle.
\end{eqnarray}
In Lyapunov control, the target state $|\psi_T\rangle$ is usually
chosen as an eigenstate of the free Hamiltonian $H_0$, such that
when completing the control, the target state should be a  steady
state,
\begin{eqnarray}\label{15}
H_0|\psi_T\rangle=\lambda_f|\psi_T\rangle,
\end{eqnarray}
where $\lambda_f$ is the eigenvalue corresponding to the eigenstate
$|\psi_T\rangle$. To design the control fields, a Lyapunov function
has to be chosen. By the merit of Lyapunov control, we choose the
Lyapunov function as follows,
\begin{eqnarray}\label{15}
V=\langle\psi|P|\psi\rangle,
\end{eqnarray}
where the hermitian operator $P$ is time independent. Furthermore,
$P$ should commute with the Hamiltonian $H_0$, i.e., $[H_0,P]=0$.
With this definition, the time derivative of $V$ is
\begin{eqnarray}\label{15}
\dot{V}=\sum_{k=1}^{K}f_k(t)\langle\psi|i[H_k,P]|\psi\rangle.
\end{eqnarray}
The construction of  the operator $P$ is flexible, such as
\begin{eqnarray}\label{15}
P=p_f|\lambda_f\rangle\langle\lambda_f|+\sum_{i\neq
f}p_i|\lambda_i\rangle\langle\lambda_i|,
\end{eqnarray}
where $|\lambda_i\rangle$ ($i=1,...,N$) is the eigenstate of
Hamiltonian $H_0$ with  eigenvalue $\lambda_i$. To transfer a
quantum state successfully,  we should choose $p_i$ to meet the
condition that $p_f<p_i$ \cite{yi09}. Obviously, when the system
arrives at the target state, $V$ reaches its minimum. Simple algebra
shows that the control fields $f_k(t)=-A_{k}\langle
\psi|i[H_k,P]|\psi\rangle$ with $A_{k}>0$ can assure $\dot{V}\leq0$.
As a consequence, those control fields would transfer the quantum
state from the spin 1 to the spin $M$.

Strictly speaking, the proposal of state transfer should work for
any state at spin 1, i.e., $\alpha$ and $\beta$ in equation
(\ref{ini1}) could take any value. A good measure to quantify the
performance of the state transfer   then should take an average over
all $\alpha$ and $\beta$. With this consideration,  we introduce the
following averaged fidelity $F(t)$ to quantify the state transfer
\cite{bose03},
\begin{eqnarray}\label{15}
F(t)=\frac{|f_{M,1}(t)|}{3}\cos\tau+\frac{|f_{M,1}(t)|^2}{6}+\frac{1}{2},
\end{eqnarray}
where $\tau=\arg|f_{M,1}(t)|$ and $f_{M,1}(t)=\langle
\textbf{M}|e^{-iHt}|\textbf{1}\rangle$. We set $\cos\tau=1$
hereafter since $\tau$ is controllable.

\section{target state}\label{III}

Conventionally, the target state should be an eigenstate of the free
Hamiltonian, such that when finishing the control, the target state
is stationary. However, the perfect goal of this proposal is to
transfer a state from one side $|\textbf{1}\rangle$ to another
$|\textbf{M}\rangle$.  To this extent, our proposal cannot
be taken as a proposal for state transfer, nevertheless, when
$|\lambda_f\rangle$ is almost the same as $| \textbf{M}\rangle$, our
proposal works. So, the purpose of this section is to choose the
parameters such that $|\langle \textbf{M}|\lambda_f\rangle|$ is as
high as possible (i.e., making the occupation of spin $M$ approach 1
in the eigenstate $|\lambda_f\rangle$). In the following, we
calculate the eigenstates of Hamiltonian $H_0$ in the spin-1/2 chain
where the Larmor frequencies $\Omega_m$ and coupling strength $D_m$
in equation (\ref{3}) are periodic function of sites with period
$l$, i.e. $\Omega_m=\Omega_{m+l}$  and $D_m=D_{m+l}$. To be
specific, we set the period $l=3$. The solutions of eigenvalues and
eigenstates of this general linear chain with open boundary have
been given in appendix \cite{feldman06}. We select a specific
eigenstate of the Hamiltonian $H_0$ so that the target state
satisfies the condition $|\langle \textbf{M}|\lambda_f\rangle| \sim
1$. By the results given in the appendix, the components of this
specific eigenstate $u$ can be expressed analytically as follow,
\begin{eqnarray}\label{133}
u_{3i-1}&=&\frac{D_1}{\lambda_{v}-\Omega_2}u_{3i-2},    \nonumber \\
u_{3i}&=&0,            \nonumber \\
u_{3i+1}&=&-\frac{D_1D_2}{(\lambda_{v}-\Omega_2)D_3}u_{3i-2},
\end{eqnarray}
where $i=1,...,n$ and the corresponding eigenvalue
$\lambda_v=\frac{1}{2}[(\Omega_1+\Omega_2)\pm\sqrt{(\Omega_1-\Omega_2)^2+4D_1^2}]$.
From the expression of eigenstate $u$, it is not hard to find that
the occupations on each site (defined as $|u_i|^2$) are determined
by the coupling strengths $D_{k} (k=1,2,3)$ and the Larmor
frequencies $\Omega_1$ and $\Omega_2$. If the conditions
$|\frac{D_1}{\lambda_{v}-\Omega_2}|<1$ and
$|\frac{D_1D_2}{(\lambda_{v}-\Omega_2)D_3}|<1$ are both satisfied
simultaneously, the values of $|u_{3i-2}|^2$ and $|u_{3i-1}|^2$
decrease monotonously, respectively. In this case, the occupation on
the spin $M$ is maximum or minimum, hence it is possible to let this
eigenstate $u$ serve as the target state. In particular,
$f(D_1)=\frac{D_1}{\lambda_{v}-\Omega_2}$ is a monotonic increasing
function with the argument $D_1$. Thus, the value of $D_1$ should be
small in order to obtain high fidelity state transfer, i.e., the
boundary spins should be weakly  coupled   to the bulk of spin-1/2
chain.

\begin{figure}[h]
\centering
\includegraphics[scale=0.5]{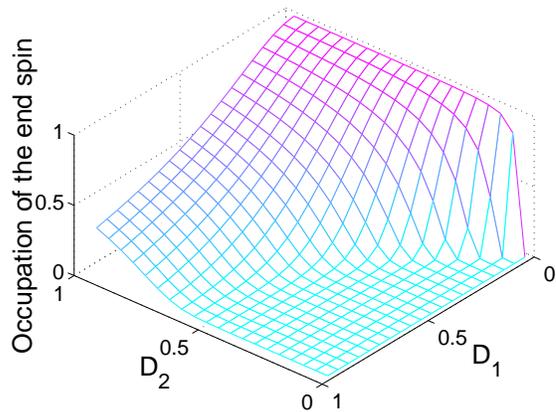}
\caption{The occupation of the end spin as a function of interaction
strength $D_1$ and $D_2$. We choose the total number of spins
$M=29$. The other parameters are in units of the
coupling strengths $D_3=1$, and $\Omega_1=1.5$,
$\Omega_2=\Omega_3=0.75$.} \label{fig:1}
\end{figure}

Instead of the quadratic couplings investigated in Ref.
\cite{bruderer12}, we here consider  periodic couplings for the
following reasons. Firstly,  the system is  robust against small
perturbations due to the existence of gaps; Secondly, it can
describe topological materials like topological insulators. In
addition, systems with periodic couplings are not rare in practice.
Figure \ref{fig:1} shows the occupation of end spin $M$ as a
function of the coupling strengths $D_1$ and $D_2$. As expected, the
occupation of end spin $M$ becomes large when $D_1$ is small. For
simplicity, we choose $D_1=0.15$, $D_2=D_3=1$, $\Omega_1=1.5$, and
$\Omega_2=\Omega_3=0.75$ in the following numerical calculation.

\section{realization of state transfer by lyapunov control}\label{IV}

\subsection{Lyapunov function}

\begin{figure}[h]
\centering
\includegraphics[scale=0.5]{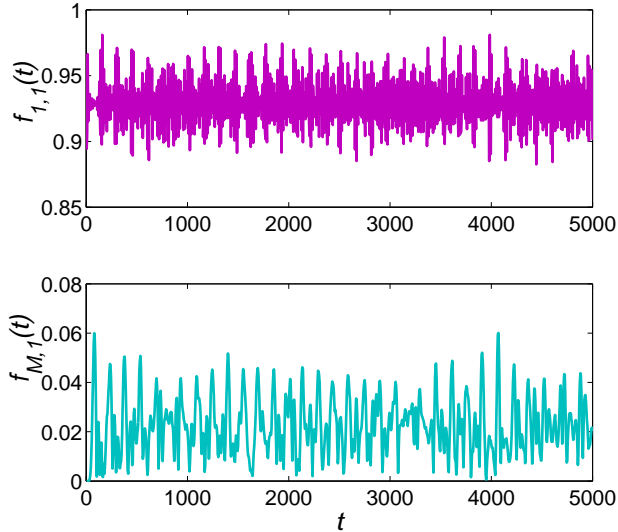}
\caption{Free evolution, namely, the time evolution without any
controls. (a) The fidelity $f_{1,1}(t)=\langle
\textbf{1}|e^{-iH_0t}|\textbf{1}\rangle$ versus   time $t$. (b) The
fidelity $f_{M,1}(t)=\langle
\textbf{M}|e^{-iH_0t}|\textbf{1}\rangle$ versus   time $t$.}
\label{fig:01}
\end{figure}

In this section, we first investigate the free   evolution of the
system starting with initial state $|\textbf{1}\rangle$ without any
external controls, that is, $H_k=0$ in equation (\ref{7}). The
results are presented in figure \ref{fig:01}. One can find   that
the state of system stays at $|\textbf{1}\rangle$ with a high
fidelity. This is a merit of our spin-1/2 chain: not only the final
state but the initial state are almost stationary when the system
does not subject to any external controls. In other words, it can
not realize state transfer without  external controls in our system
since the system is almost still at the initial state.

\begin{figure}[h]
\centering
\includegraphics[scale=0.5]{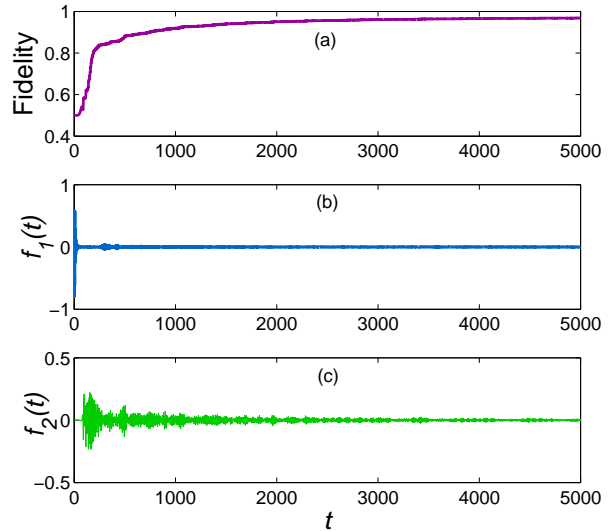}
\caption{(a) Evolution of the fidelity $F$ versus control time $t$.
We have set $A_{k}=1$ and $P_f=-3$ to guarantee   the condition
$P_f<P_i$ with  $P_i=\lambda_i$. (b) The   control field $f_1(t)$.
(c) The control field $f_2(t)$.} \label{fig:2}
\end{figure}

Next, we control the coupling strengths  between the boundary spins
and the bulk spins, i.e.,
\begin{eqnarray}\label{13}
H_{k}=c_{k}^{\dag}c_{k+1}+c_{k+1}^{\dag}c_{k},~~ k=1,M-1.
\end{eqnarray}
The coefficients of the hermitian operator $P$ are chosen as
$p_i=\lambda_i$ and $p_f=-3$. The dynamics behavior of fidelity $F$
is illustrated as a function of the control time $t$ in figure
\ref{fig:2}. It approaches a steady value about 96.84\% when the
control finishes, thus we can conclude that we have realized state
transfer with high fidelity.

\subsection{ Robustness of state transfer}
So far, we have demonstrated that  high fidelity of state transfer
can be obtained by Lyapunov control  without any fluctuations in the
control field and uncertainties in the free Hamiltonian $H_0$. In
practice, however, the parameters in the free Hamiltonian $H_0$ are
not easy to acquire precisely, and fluctuations of the control
fields might occur in experimental manipulations, which would lead
to errors in the state transmission. In this section, we focus on
this issue. We begin with analyzing robustness against the
uncertainty in the free Hamiltonian $H_0$, which can be represented
by $\delta\cdot H_0$,
\begin{eqnarray}\label{15}
H_0\rightarrow H_0+\delta\cdot H_0.   \nonumber
\end{eqnarray}
Here $\delta$ is independent random quantities manifested in
coupling strengths $\delta\cdot D_k$ or Larmor frequencies
$\delta\cdot \Omega_k$ ($k=1,2,3$). Since the fidelity of this
system will be impacted by the perturbations seriously  when the
control time takes too long, we truncate the control time $t=1000$
in the numerical calculations while the fidelity can  approach 92\%.
As we only control the boundary coupling strengths to achieve state
transfer, we emphasize on analyzing the influence of boundary
parametric fluctuations on the fidelity at first.

\begin{figure}[h]
\centering
\includegraphics[scale=0.5]{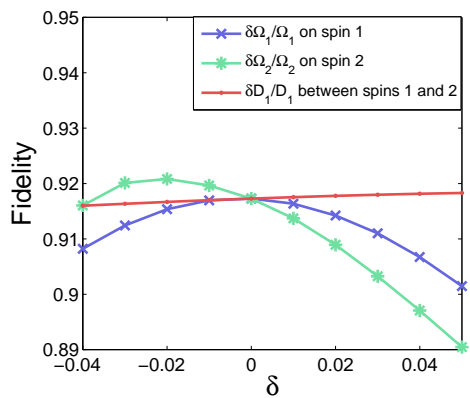}
\caption{The behavior of fidelity versus disorder of parameters
related to spin 1 and spin 2. We have fixed the control time
$t=1000$.} \label{fig:3}
\end{figure}

\begin{figure}[h]
\centering
\includegraphics[scale=0.5]{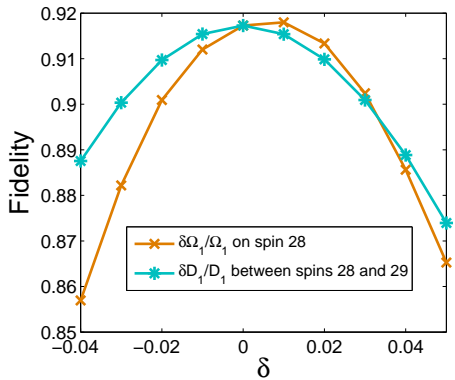}
\caption{The behavior of fidelity versus disorder of the magnetic
field $\Omega_1$ on spin 28 and the coupling strength between spins
28 and 29. We have fixed the control time $t=1000$.} \label{fig:4}
\end{figure}

\begin{figure}[h]
\centering
\includegraphics[scale=0.5]{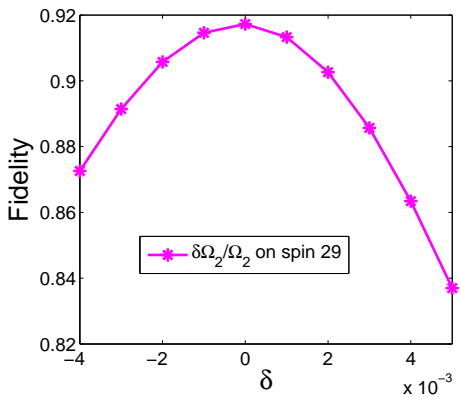}
\caption{The behavior of fidelity versus disorder of the magnetic
field $\Omega_2$ on spin 29. We have fixed the control time
$t=1000$.} \label{fig:5}
\end{figure}

\begin{figure}[h]
\centering
\includegraphics[scale=0.5]{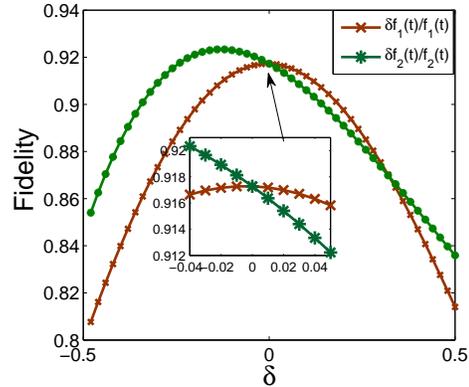}
\caption{The behavior of fidelity versus disorder of the control
fields. We have fixed the control time $t=1000$.} \label{fig:6}
\end{figure}

Figure \ref{fig:3} shows the fidelity of state transfer as a
function of uncertainties in  the first or second spins. Those
perturbations have slight effects on the fidelity of state transfer
since those parameters primarily affect the control field $f_1(t)$,
and it only plays a significant role in the beginning while almost
vanishes at later time, as seen in figure \ref{fig:2}. Figure
\ref{fig:4} and figure \ref{fig:5} show the relationship between the
fidelity and the uncertainties  related to the $28$th or $29$th
spins. Those parameters have much impact on the fidelity due to the
closely connection with the control field $f_2(t)$ and the
occupation of end spin in target state. In addition, it can be found
in those figures that the fidelity might be even   higher when the
uncertainties are very small. The reason can be understood as
follows. In the ideal situation, the fidelity increases monotonously
during the time evolution. Small uncertainties/fluctuations  might
not change the sign but change the amplitude of control fields,
leading to a slight oscillation in the value of fidelity. As a
consequence, the fidelity might increase due to these uncertainties.

This can also be used to explain why   a higher fidelity can be
reached when fluctuations  exist  in the control fields $f_k(t)$,
since the fidelity sharply depends on the sign rather than the
amplitude of control fields in figure \ref{fig:6}. We find from the
inset that the influence of the control field $f_2(t)$ is more
sensitive than that of the control field $f_1(t)$ when fluctuations
are  small. It can be explained by an observation in figure
\ref{fig:2} that there still exists small control field $f_2(t)$
when $t>1000$ while the amplitude of control field $f_1(t)$ almost
vanishes at that time. For its insensitive in the fluctuations of
control fields, making the interaction time long would be in favor
of obtaining high fidelity of state transfer. On the other hand,
when considering the disorder of other physical parameters such as
coupling strengths $\delta\cdot D_k$ or Larmor frequency
$\delta\cdot \Omega_k$, the distribution of eigenstates of the
Hamiltonian $H_0$ might change. Hence the final state might not be a
steady state and evolves even though the control fields disappear
(or turn off). So the fidelity of state transfer might deteriorate
with the increasing of interaction time. As a consequence, one
should trade off the disorders of different parameters to truncate
an appropriate control time in order to obtain a relatively high
fidelity.
\begin{figure}[h]
\includegraphics[scale=0.5]{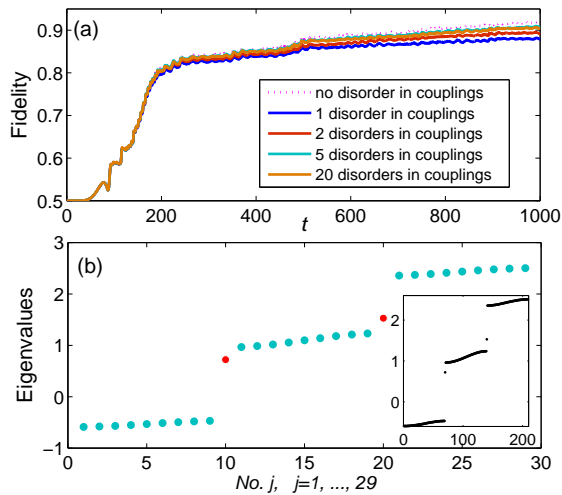}
\caption{(Color online) (a)Fidelity as a function of time.  The dot
line denotes the fidelity in the case  without  disorders,  while
the solid   lines show the fidelity with disorders. Each line is an
average over 100 simulations, where each simulation is obtain by
introducing no disorder, 1, 2, 5 and 20 disorders.  (b)The
eigenvalues of Hamiltonian when the total spins $M=29$. The inset
shows the eigenvalues of total spins $M=209$. } \label{BBfig}
\end{figure}

Now, we turn to study the effect of disorder in (the whole
spin chain) couplings on the performance of the state transfer. To
this end, we adopt the same control fields and initial state as in
the case without disorders, but change randomly the couplings from
$D_j$ to $D_{j}^{r}=(1+\epsilon)D_{j}$, where $D_{j}$ is the
original couplings in the above analysis and $j$ is spin site index.
$\epsilon$ is a  randomly chosen number in the interval [-0.05,
0.05]. In other words, we explore the effect of disorders by
re-simulating the state transfer with $D_j^r$ instead of $D_j$, and
$D_j^r$ changes randomly at each point of time for a randomly chosen
spin site. We simulate $n$ disorders ($n=1,2,5,20$) existing
simultaneously in the system and show the results in figure
\ref{BBfig}(a). One can discover that the scheme is robust against
disorders in the couplings. The disorders in the on-site energy
would have similar effects on the performance. 

An interesting observations is that more disorders might
benefit the performance of the state transfer. This might relate to
the random number $\epsilon$ being  created in [-0.05, 0.05], i.e.,
the average of $\epsilon$ is closely to zero. The physics behind the
robustness of the scheme is that  small random disorders in the
couplings cannot close the gaps  in the system since it has
nontrivial topological property \cite{lang12}, as shown in figure
\ref{BBfig}(b). As a result,  the boundary states appear in the
system, making our control protocol robust, which is the essential
reason why we choose the periodical system.

\section{scalability and discussion}\label{V}

\begin{figure}[h]
\centering
\includegraphics[scale=0.5]{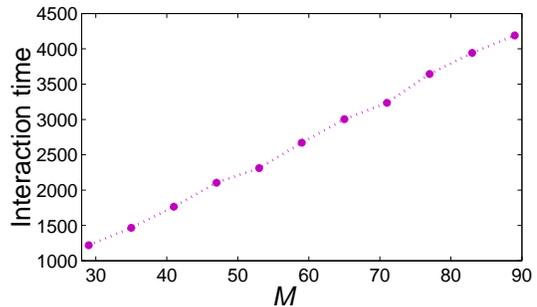}
\caption{ The interaction time $t$ versus different total number of
spins $M$. The interaction time finishes when the fidelity of state
transfer reaches 0.93. All the other parameters are the same with
figure \ref{fig:2}.} \label{fig:7}
\end{figure}

Until now, we have just considered a chain with $M=29$ spins. A
natural question arises  that whether the interaction time increases
with the increasing of the spin number. Since the interaction time
is also closely related to the amplitude of control fields $f_k(t)$,
we set $A_k=1$ in the expression of $f_k(t)$ and study this problem
by numerical calculation. Figure \ref{fig:7} demonstrates the
relation between the interaction time and different total number of
spins while the fidelity of state transfer reaches 0.93. It is
observed that the interaction time increases approximately linearly
with the total number of spins.

\begin{figure}[h]
\centering
\includegraphics[scale=0.5]{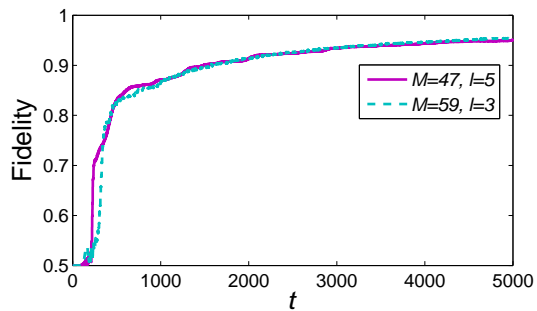}
\caption{ Evolution of the fidelity $F$ versus control time $t$. The
parameters (the pink solid line) are $\Omega_1=\Omega_5=1.5$,
$\Omega_2=\Omega_3=\Omega_4=0.75$, $D_1=0.15$, $D_3=0.5$,
$D_2=D_4=D_5=1$. The parameters of the cyan dash line are the same
to figure \ref{fig:2} except for the total number of spins $M=59$.}
\label{fig:8}
\end{figure}

Although the general formulas of exact diagonalization of matrix
with  period ($l>3$) in Larmor frequencies and coupling strengths is
given in Ref. \cite{feldman06}, the analytical expressions are
complicated. For this reason, we will employ the numerical
calculation to investigate whether the Lyapunov control is suited to
the case for the period $l>3$. Numerical simulations show that, it
is possible to obtain high fidelity state transfer in a  spin-1/2
chain with 47 sites of period $l=5$, and the fidelity  can approach
94.97\% when completing the control, as shown in figure \ref{fig:8}.
The another feature in figure \ref{fig:8} is that the dynamics
behaviors of the fidelity with $M=47$ and $l=5$ almost coincides
with the case of $M=59$ and $l=3$, indicating that the interaction
time does not linearly increase  with the total number of spins in
different periodic  structure of spin-1/2 chains. As the Lyapunov
function, i.e., $V=p_f|\langle\psi|\lambda_f\rangle|^2+\sum
p_i|\langle\psi|\lambda_i\rangle|^2$, represents the weighted
average between the state of system and distinct eigenstates of
Hamiltonian $H_0$. The variation of Lyapunov function reflects that
the control fields mainly modulate the state transition between the
eigenstates of Hamiltonian $H_0$. Thus the interaction time is
connected with the characteristic spectrum of Hamiltonian $H_0$
rather than the total number of spins. Once the free Hamiltonian
$H_0$ of spin-1/2 chain contains an eigenstate whose occupation
mainly locates at the end spin, there should exist a Lyapunov
function  for the control fields to control the state transition
between eigenstates of Hamiltonian $H_0$. As a result, it can
realize state transfer by the Lyapnouv control, of course, the
fidelity of state transmission must be closely linked to the
characteristic spectrum of Hamiltonian $H_0$.

\begin{figure}[h]
\centering
\includegraphics[scale=0.5]{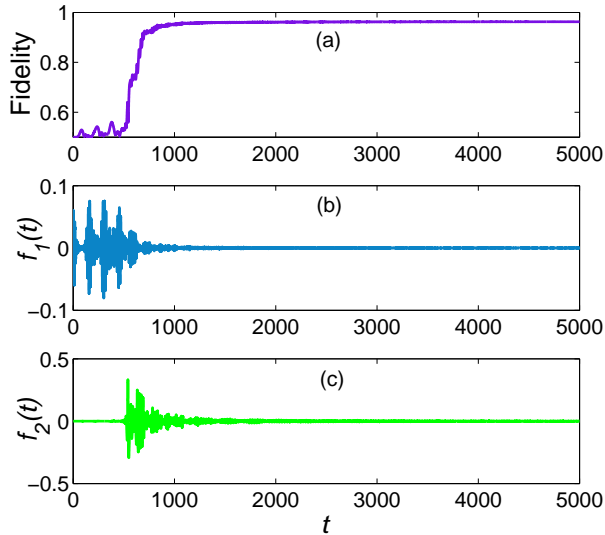}
\caption{All  parameters and the expression for control function are
the same as in figure \ref{fig:2} except for the control
Hamiltonian.} \label{fig:02}
\end{figure}

\begin{figure}[h]
\centering
\includegraphics[scale=0.5]{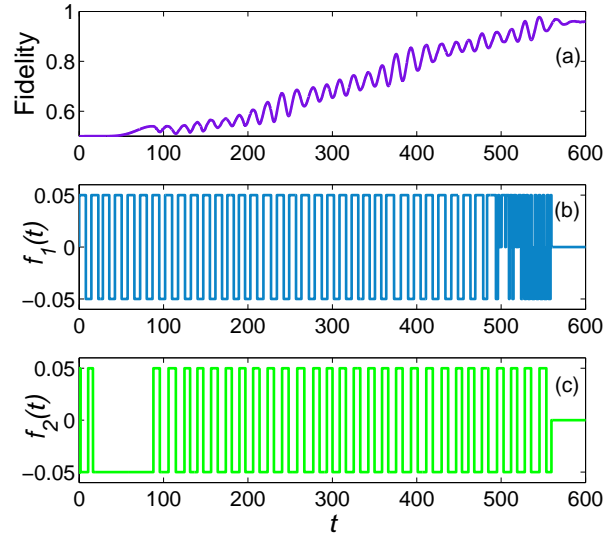}
\caption{All physical parameters are the same as in figure
\ref{fig:02} except for the control fields. We set
$F^{\prime}_k=0.05$. The control  is terminated  when the fidelity
arrived at 0.96.} \label{fig:03}
\end{figure}

In a practical situation, the control Hamiltonians which exchange
the interaction between the boundary spins and its neighbors in
equation (\ref{13}) might be difficult to obtain. An alternative way
of realizing state transfer is to control the Larmor frequencies at
the boundary spins, namely, the control Hamiltonians are chosen as
$H_1=c^{\dag}_1c_1$ and $H_2=c^{\dag}_Mc_M$. Figure \ref{fig:02}
shows the performance of fidelity when controlling the Larmor
frequencies at the boundary spins. Indeed, it also can realize state
transfer. Furthermore, we can replace the fast time-varying control
fields by square wave pulses, which reduces the difficulty in
experiments. The square wave pulses are  given by
\begin{eqnarray}\label{21}
f_k(t)=\left \{
\begin{array}{rl}
    F^{\prime}_k,~~~ f_k(t)>0, \\
    -F^{\prime}_k,~~~f_k(t)<0. \\
\end{array}
\right.
\end{eqnarray}
By these square wave pulses, in figure \ref{fig:03}, we show that
the target can be reached with high fidelity as well.

\section{conclusion}\label{VI}

In conclusion, by Lyapunov control technique, we have proposed a
scheme for state transfer in the spin-1/2 chain. The controls are
exerted on the couplings between the boundary spins and its
neighbors. It also works by  manipulating  the Larmor frequencies of
the boundary spins. The difference between our proposal and
the others  is that the resulting state here is an eigenstate of the
system Hamiltonian, hence it is stationary and it does not require
to finish the transfer at a fixed time.  In addition, the proposal
is robust against many types of perturbations and disorders
since those perturbations could not close the gaps of the
system.  This proposal  can also  be generalized to a structure
with periodicity $l>3$, provided that the chain has an eigenstate
localized at the boundary spin (the receiver). For chains which have
no eigenstate perfectly localized  at the boundary spin, a high
fidelity state transfer can be obtained as well by elaborately
designing the parameters of the spin-1/2 chain. Finally, we have
shown that replacing the fast time-varying control fields with
square wave pulses are possible, which simplifies the experimental
realization.

\section*{ACKNOWLEDGMENTS}

This work is supported by the National Natural Science Foundation of China (Grants No. 11175032 and No.
61475033).

\section*{APPENDIX}
The diagonalization  of a matrix representing the Hamiltonian of
spin-1/2 chain with  total spins $M$ ($M=nl+d$), Larmor frequencies
$\Omega_m$, coupling strength $D_m$,  periodicity  $l$, $l>d\geq0$
and $l>1$, has been given  in Ref.
\cite{feldman06,schultz69,feldman05}. When $l=3$ and $d=2$, the
matrix reads,
\begin{equation}
{H}=\left( \begin{array}{ccccccc}
   \Omega_1 & D_1 & 0 & 0 & \cdots & 0 & 0  \\
   D_1 & \Omega_2 & D_2 & 0 & \cdots & 0 & 0  \\
   0 & D_2 & \Omega_3 & D_3 & \cdots & 0 & 0 \\
   0 & 0 & D_3 & \Omega_1 & \cdots & 0 & 0 \\
   \vdots & \vdots & \vdots & \vdots & \ddots & \vdots & \vdots \\
   0 & 0 & 0 & 0 & \cdots & \Omega_1 & D_1 \\
   0 & 0 & 0 & 0 & \cdots & D_1 & \Omega_2 \\
\end{array} \right).
\end{equation}
The  $3n$-distinct eigenvalues $\lambda_v$ of this Hamiltonian can
be obtained by solving following equation,
\begin{eqnarray}\label{15}
(\lambda_{v}-\Omega_1)(\lambda_{v}-\Omega_2)(\lambda_{v}-\Omega_3)-
(\lambda_{v}-\Omega_2)D_3^2     \nonumber\\
-(\lambda_{v}-\Omega_1)D_2^2-(\lambda_{v}-\Omega_3)D_1^2=2D_1D_2D_3\cos(\frac{\pi
k}{n+1}),
\end{eqnarray}
where $k=1,2,...,n.$ The corresponding eigenvectors
$u_j=(u_{1,j},u_{2,j},u_{3,j},...,u_{3n+1,j},u_{3n+2,j})^T$ are
\begin{eqnarray}\label{15}
U_{1,j}&=&\frac{(\lambda_{v}-\Omega_2)\mathscr{D}_3^T+D_1\mathscr{D}_2^T}
{(\lambda_{v}-\Omega_1)(\lambda_{v}-\Omega_2)-D_1^2}\cdot U_{3,j},     \nonumber\\
U_{2,j}&=&\frac{(\lambda_{v}-\Omega_1)\mathscr{D}_2^T+D_1\mathscr{D}_3^T}
{(\lambda_{v}-\Omega_1)(\lambda_{v}-\Omega_2)-D_1^2}\cdot U_{3,j},     \nonumber\\
U_{3,j}&=&\Big(\sin(\frac{\pi [\frac{j+2}{3}]}{n+1}),
\sin(\frac{2\pi [\frac{j+2}{3}]}{n+1}), ... , \sin(\frac{n\pi
[\frac{j+2}{3}]}{n+1})\Big)^T \nonumber
\end{eqnarray}
with
\begin{eqnarray}\label{15}
U_{1,j}&=&(u_{1,j},u_{4,j},...,u_{3n-2,j},u_{3n+1,j})^T,    \nonumber\\
U_{2,j}&=&(u_{2,j},u_{5,j},...,u_{3n-1,j},u_{3n+2,j})^T,    \nonumber\\
U_{3,j}&=&(u_{3,j},u_{6,j},...,u_{3n,j})^T,     \nonumber
\end{eqnarray}
\begin{equation}
{\mathscr{D}_2}=\left( \begin{array}{ccccc}
   D_2 & 0 & \cdots & 0 & 0 \\
   0 & D_2 & \cdots & 0 & 0 \\
   \vdots & \vdots & \ddots & \vdots & \vdots \\
   0 & 0 & \cdots & D_2 & 0 \\
\end{array} \right),       \nonumber
\end{equation}
\begin{equation}
{\mathscr{D}_3}=\left( \begin{array}{ccccc}
   0 & D_3 & 0 & \cdots & 0 \\
   0 & 0 & D_3 & \cdots & 0 \\
   \vdots & \vdots & \vdots & \ddots & \vdots \\
   0 & 0 & 0 & \cdots & D_3            \nonumber
\end{array} \right).
\end{equation}
Here $[x]$ represents the largest integer less than or equal to $x$,
the superscript ``$T$'' denotes transposition and $j=1,2,...,3n$.
$\mathscr{D}_2$ and $\mathscr{D}_3$ are $n\times (n+1)$ matrices.

The other two eigenvalues $\lambda_v$ satisfy the following
equation,
\begin{eqnarray}\label{15}
(\lambda_{v}-\Omega_1)(\lambda_{v}-\Omega_2)-D_1^2=0.
\end{eqnarray}
The corresponding eigenvector $u_j$ can be expressed
\begin{eqnarray}\label{15}
u_{3i-1,j}&=&\frac{D_1}{\lambda_{v}-\Omega_2}u_{3i-2,j},    \nonumber\\
u_{3i,j}&=&0,           \nonumber\\
u_{3i+1,j}&=&-\frac{D_1D_2}{(\lambda_{v}-\Omega_2)D_3}u_{3i-2,j},
\nonumber
\end{eqnarray}
where $i=1,...,n$ and $j=3n+1,3n+2$.


\begin{references}
\bibitem{note1} The state $|0\rangle$ ($|1\rangle$) is equivalent to
spin down (up) in spin-1/2 chain.
\end{references}
\end{document}